# Social Media Data Analysis and Feedback for Advanced Disaster Risk Management

Markus Enenkel
International Research Institute for Climate and Society
The Earth Institute at Columbia University, NY 10964, USA.
Global Emergency Response Team, SOS Children's Villages International, 1200 Vienna, Austria
menenkel@iri.columbia.edu

Sofía Martinez Sáenz
International Research Institute for Climate and Society
The Earth Institute at Columbia University, NY 10964, USA.
sms@iri.columbia.edu

Denyse S. Dookie
School of International and Public Affairs (SIPA), Columbia University, New York, 10027, USA
dsd2123@columbia.edu

Lisette Braman
International Research Institute for Climate and Society
The Earth Institute at Columbia University, NY 10964, USA
lisette@iri.columbia.edu

Nick Obradovich
Scalable Cooperation Group, Media Lab, Massachusetts Institute of Technology, Cambridge, MA 02139 USA
nobradov@mit.edu

Yury Kryvasheyeu
Data61, Commonwealth Scientific and Industrial Research Organisation (CSIRO), Docklands 3008, Victoria, Australia
Faculty of Information Technology, Monash University, Caulfield 3145, Victoria, Australia
yury.kryvasheyeu@data61.csiro.au

## ABSTRACT
Social media are more than just a one-way communication channel. Data can be collected, analyzed and contextualized to support disaster risk management. However, disaster management agencies typically use such added-value information to support only their own decisions. A feedback loop between contextualized information and data suppliers would result in various advantages. First, it could facilitate the near real-time communication of early warnings derived from social media, linked to other sources of information. Second, it could support the staff of aid organizations during response operations. Based on the example of Hurricanes Harvey and Irma we show how filtered, geolocated Tweets can be used for rapid damage assessment. We claim that the next generation of big data analyses will have to generate actionable information resulting from the application of advanced analytical techniques. These applications could include the provision of social media-based training data for algorithms designed to forecast actual cyclone impacts or new socio-economic validation metrics for seasonal climate forecasts.

## 1. INTRODUCTION
The damages caused by recent hurricanes Harvey, Irma, Maria, and Nate, serve as a premonition of predicted future climatic changes [1, 11]. Together these disasters resulted in an estimated US$477.5 billion in damages and 198 official deaths for the United States alone [3]. Emanuel [4] notes that while there has been a steady improvement in hurricane track forecasts, the prediction of hurricane intensity and related damages remains challenging. Some of the storms in the 2017 season (most notably Harvey as it approached Texas, and Maria as it affected Dominica) went through rapid intensification just prior to landfall, an occurrence which tends to happen 10-20 times more frequently in a warmer atmosphere [4]. There is an urgency to systematically understand and communicate such risks and decision options in timely and actionable ways.

In addition, near real-time response to such information, appreciating the socioeconomic and physical needs of those requiring assistance, is critical. One of the major limitations from a humanitarian perspective is that aid organizations need to cope with high uncertainties regarding where to conduct in-depth damage assessments. Any information that directs aid workers' assessment teams is considered helpful from a practical point of view. While there are efforts to "connect the dots" based on the exploitation of big data, there remains an inherent need to ensure that information to and from the affected public is well-understood and well-utilized. The technologies to translate big data into actionable information, as well as tools to communicate information related to risk and decision-support to users in near real-time, exists. These are likely feasible solutions to transform social media users from exclusively passive data suppliers to active agents of information, and to effectively utilize social media as a two-way communication channel.

This position paper explores a potential solution to the data, information and knowledge management challenges in natural disaster risk management. It offers an improvement to the existing uses of mainstream social media to assist in harnessing, digesting, disseminating and dynamically updating decision options. Social media data has been recently gaining momentum as a useful indicator for needs and damage assessment. However, we suggest that it can still encourage and enable near real-time two-way communication and response, both in the immediate and long-term perspectives of disaster risk management. In line with this practical approach, we present an example of plausible damage distribution maps for hurricanes Harvey and Irma in the US context. We suggest these maps could be used to inform disaster managers' view of existing risk, helping identify and validate real needs, resulting in a targeted and timely response. Such courses of action could certainly complement ongoing disaster preparedness efforts, and likely amplify the humanitarian response in the post-disaster context through more efficient resource allocation.

## 2. USING SOCIAL MEDIA IN A DISASTER RISK MANAGEMENT CONTEXT

Crowdsourcing and various applications of social media data in the context of disaster management have been studied extensively since the early days of online social media platforms. There is a solid understanding of emergency information diffusion, role of social media platforms in gathering and disseminating news, their contribution to situational awareness, and adoption by formal respondents to serve public demand for crisis-related information [7, 13, 15, 16]. This accumulated knowledge begins to translate into applied tools that are important for effective disaster risk management [2, 5, 14]. For instance, the Humanitarian OpenStreetMap Team relies on large groups of private citizens to map the condition of critical infrastructure via high resolution satellite imagery after disasters. Facebook introduced a feature that allowed people in emergencies to tag themselves as "safe".

However, there are still important gaps related to the use of data gathered from social media channels when it comes to actual decision-support. The US Federal Emergency Management Agency (FEMA) considers social media "an important part of the whole community approach because it helps facilitate the vital two-way communication between emergency management agencies and the public."[1] However, there are two interrelated issues. First, the "two-way" component remains limited to internal decision-support, not to the direct interaction with people in the field to identify needs, forward early warnings, assess damages and target response efforts. The FEMA app[2] allows users to upload photos, to receive early warnings and information about shelters. Also Facebook is working on disaster maps to provide information about the location of populations, their movements and if they're "checking in safe during a natural disaster."[3] However, to our knowledge, there is no analytical link between data supply and provision (feedback). Second, although the potential of unintentionally provided and crowdsourced data (e.g. geolocated tweets) for rapid damage assessments is demonstrated for a range of disasters [9], it remains largely untapped.

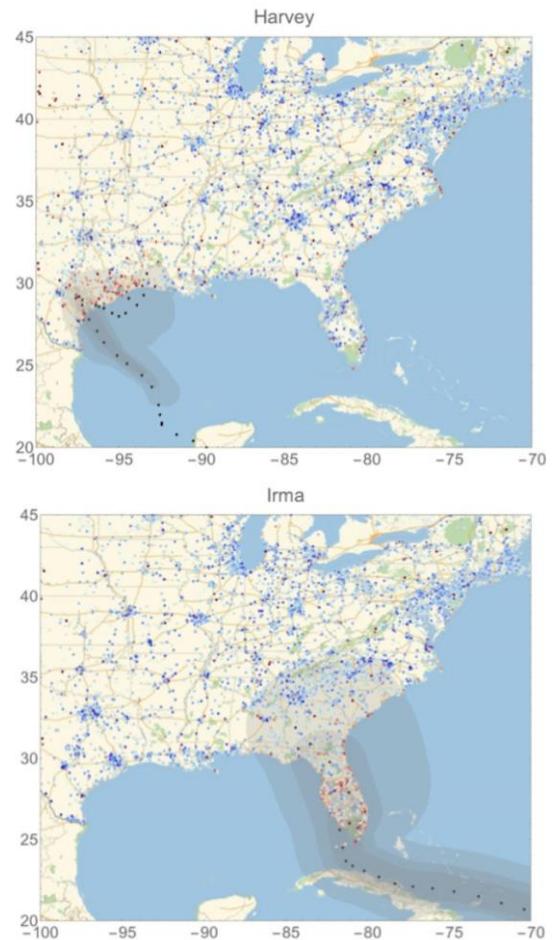

**Figure 1.** Spatial distribution of hurricane-related tweets approximates the damage inflicted by Hurricane Harvey (top) and Hurricane Irma (bottom).

Tapping into this rich database can allow us to estimate the spatial distribution of damages virtually at no cost, and in a timely manner. **Figure 1**, above, highlights the spatial distribution of damage estimates based on population-adjusted density of disaster-related tweets for recent hurricanes Harvey and Irma. The heatmap represents the density of disaster-related messages. In our spatial analysis we rely on the boundaries and population estimates provided by the U.S. Census Bureau (ZIP code tabulation areas, or ZCTAs). Each point in the figure represents the centroid of a corresponding area, and the color reflects the per-capita number of disaster-related messages posted within it, varying from low density (blue) to high density (red) of tweets. High per-capita count of social media messages related to a disaster is shown to correlate with high damage [9] and the map could be directly interpreted as preliminary estimate of damage distribution.

Such estimates can be easily generated by continuous monitoring of Twitter stream for a set of predefined keywords. Since the correlation between online activity and damage is strongest in the post-disaster phase [9] we analyze tweets after the peak (hurricane landfall) of each event. Rapid in nature, and improving in precision as the data collection runs continuously, such estimates may inform and guide response and relief operations.

---

[1] See https://emilms.fema.gov/is42/BPSM0102010t.htm
[2] See https://www.fema.gov/mobile-app
[3] See https://research.fb.com/facebook-disaster-maps-methodology/.

Understanding and utilizing this data source of real-time damage assessment and estimates, produced by local community members who may be actively affected by the event, could likely assist in quicker response and recovery efforts. This is since early warnings with a physical dimension approach (e.g. cyclone trajectories) are very limited when it comes to humanitarian decision-making (preparedness), without recent information about local vulnerabilities, coping capacities (as they relate to available resources, both human and capital) and the translation of early warnings into early action. In addition, for some disasters like flash floods or earthquakes, not to mention technological catastrophes, the lead times for early warnings are very short or non-existent, potentially turning social media into sources of early warning information themselves. Since the same tools and apps that are used for status updates in social media can be used to collect socio-economic information at high temporal frequencies, we argue it is also possible to develop a new generation of baseline datasets and trend indicators. While active data provision via surveys requires incentives, the analysis of passively supplied data like call detail records can result in big advantages for operational decision-making, for instance via the analysis of migration streams [10].

Moreover, we encourage the development of algorithms to understand the community needs as suggested in these tweets to directly target and attend to the specific challenges in various localities given the known financial, human and physical vulnerabilities. Live-responding to and directly following-up on tweets could foster optimal two-way communications between those in need and disaster agencies. This approach could improve the assessment of the problem and potential gaps in response, allowing community-based solutions where possible and directing resources in a more timely and efficient manner.

Several research projects discovered the added-value of social media and big data for disaster risk management. Also scientific studies about the potential bias towards higher-income regions is available [e.g. 8, 11]. However due to the focus on geographic bias in high-income countries (e.g. the United States) these studies do not answer the question of which group of social media users acts as proxy for the general population. Although, in particular, the growth rates of mobile subscriptions in low-income countries are increasing [6], there is still a wide gap between urban and rural areas. Only if we manage to understand the impact of such disparities on social media analytics will we be able to equally integrate and adequately tailor preparedness and response information to low, middle and high-income communities.

In general, we consider efficient disaster risk management to be about understanding user needs, developing user-driven solutions, holistically understanding the risk profiles and context, as well as communicating vital and actionable information. However, data collected via social media is not automatically actionable information. It needs to be analyzed along with additional datasets, such as existing survey, satellite or modeled data, to translate it into useable information and communicated to users in the field. This will require a paradigm change from individual research projects to interdisciplinary strategic development to address end-user time-sensitive needs.

## 3. THE WAY FORWARD

The next generation of social media based assessments proposed in this position paper can be regarded as a new strategy to extract actionable insights during the disaster preparedness and response phase.

While the preparedness phase can benefit from big data processing with regard to the analysis of keywords for early warning, the response phase focuses on rapid damage assessments or the justification of emergency funds allocation. However, both of these applications require a direct feedback from beneficiaries in the field to translate big data into big information via advanced filtering and validation techniques.

There is a clear need to think beyond current big data analyses to integrate actionable information derived from social media into operational decision-making. This will require humanitarian organizations to build trust on new sources of information and services that can be sourced from social media analysis. It calls for novel structures in humanitarian assistance processes, which have mostly been centralized in high-income countries (e.g. FEMA), but relies heavily on decentralized risk management in low-income countries. A promising strategy for a research-based big data solution could be based on the following four pillars:

1. Traditional standardized multi-hazard modelling tools, such as the HAZUS employed by FEMA;
2. Complementary near real-time "big-data" assessments related to early warnings, disaster trends and related impacts;
3. The integration of socio-economic and demographic data to investigate disparities related to disaster resilience; and
4. Operational feedback loops that link active data suppliers back into the decision-support system as passive users of information.

This would mean the development of an operational platform that feeds on social media data, predicts the likelihood of certain damages and intelligently takes other inputs (e.g. call detail records, census data, existing disaster databases) to learn and calibrate itself. Ideally, this platform would be sensitive to disaster types and local conditions/demographics to support decision-making for disasters that allow very little to no lead time (e.g. earthquakes). A possible logical sequence of analytical/operational development steps is outlined in **Figure 2**.

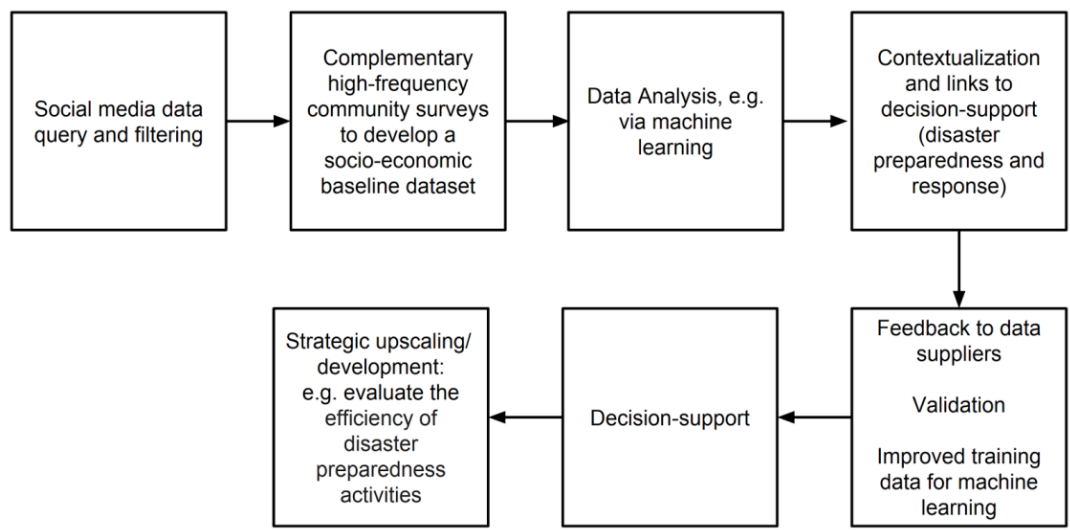

**Figure 2.** Possible logical sequence of analytical/operational development steps.

## 4. CONCLUSION

This year's *Social Web in Emergency and Disaster Management* Conference theme "Collective Sensing, Trust, and Resilience in Global Crises" indirectly highlights the role of big data to understand globalized problems. However, despite the availability of mobile applications, social media platforms, big data processing infrastructure and algorithms, the transformation of big data into actionable big information happens very slowly. On the one hand, advanced pattern detection techniques require training data, which cannot be acquired from conventional data collection strategies. The largest global socioeconomic data collection activities (e.g. USAID's Demographic and Health Surveys Program or UNICEF's Multiple Indicator Cluster Surveys) are not harmonized with each other, carried out irregularly and at large temporal intervals. On the other hand, even if these data were available to detect patterns and issue early warnings, social media and related mobile applications are mainly considered one-way streets for communication, leaving (partly unintentional) the data suppliers without an added value of their data contribution.

From a humanitarian or general disaster risk management perspective, the direction of teams on the ground to conduct rapid damage assessments, for instance after tropical hurricanes, is a major priority. Based on the example of the recent Hurricanes Irma and Harvey we show the spatial distribution of hurricane-related tweets to approximate damages. The map in **Figure 1** shows the population adjusted density of disaster-related messages. The development of an operational system that analyses various geo-located social media posts is feasible both in a disaster preparedness (e.g. to prepare for the influx of refugees) and emergency response contexts, whereas the correlation between online activity and damage is usually strongest in the post-disaster phase [9].

If coupled with other existing datasets, such a system, platform or service could serve as a complementary low-cost decision-support strategy. In a second development step, the contextualization of social media data could not only support helpers, but the affected population via conventional feedback based communications through social media. In the case a disaster affects the communication network, it can also be accomplished through peer-to-peer networking technology (e.g. Fire Chat). This step would represent the actual transformation of big data into actionable, big information via near real-time two-way communication.

The advanced use of social media analysis and two-way communication will very likely also have indirect advantages for disaster risk management. Some examples:
- It can allow the assessment of progress related to the sustainable development goals as well as the efficiency of activities aiming at increased disaster resilience, allowing to justify additional expenses for disaster preparedness.
- Social media-based damage assessments can serve as an additional training dataset for machine learning algorithms that aim at forecasting actual cyclone impacts. If successful at a usable spatial scale, this can be considered a milestone in humanitarian aid.
- The next generation of validation techniques for seasonal climate forecasts might consider alternative validation techniques. Instead of validating forecasts via actual observations of the same parameter (e.g. rainfall) it would be possible to develop new metrics representing their "socioeconomic skill".

Finally, the next generation of social media analyses will have to consider connectivity or general data gaps in the world's low-income regions as well as alternative ways of offline communication whenever disasters affect internet access (e.g. peer-to-peer networking technology).